\documentclass{article}

\def\giorno{27 February 2001}

\def\.#1{\dot #1}

\def\F{{\cal F}}
\def\G{{\cal G}}
\def\h{{\cal H}}

\def\I{{\cal I}}
\def\L{{\cal L}}
\def\M{{\cal M}}

\def\R{{\bf R}}  

\def\V{{\cal V}}
\def\W{{\cal W}}
\def\X{{\cal X}}
\def\Y{{\cal Y}} 
\def\Z{{\cal Z}} 

\def\ss{\subset}
\def \pa{\partial}

\def\=#1{\widetilde #1}
\def\~#1{\widetilde #1}
\def\.#1{\dot #1}
\def\^#1{\widehat #1}

\def\eb{{\bf e}}

\def\vb{{\bf v}}

\def\ker{{\rm Ker}}
\def\ran{{\rm Ran}}

\def\({\left(}
\def\){\right)}
\def\[{\left[}
\def\]{\right]}

\def\a{\alpha}
\def\b{\beta}

\def\phi{\varphi}
\def\la{\lambda}
\def\s{\sigma}

\def\phi{\varphi}
\def\Ga{\Gamma}

\begin{document}

\title{{\bf Algorithmic reduction of Poincar\'e-Dulac  
normal forms and Lie algebraic structure}}

\author{Giuseppe Gaeta\footnote{Work supported by ``Fondazione CARIPLO per la ricerca scientifica'' under the project ``Teoria delle perturbazioni per sistemi con simmetria''; e-mail: {\it gaeta@berlioz.mat.unimi.it} or {\it gaeta@roma1.infn.it}} \\ {\it Dipartimento di Matematica, Universit\`a di Milano} \\ {\it via Saldini 50, I--20133 Milano (Italy)} }
 
\date{\giorno}
\maketitle

\noindent{\bf Summary.} The Poincar\'e-Dulac normal form of a given  resonant system is in general non unique; given a specific normal form, one would like to further reduce it to a simplest normal form. In this note we give an algorithm, based on the Lie algebraic structure of the set of normal forms, to obtain this. The algorithm can be applied under some condition, non generic but often met in applications; when applicable, it only requires to solve linear equations, and is more powerful than the one proposed in previous 
work by the same author [{\it Lett. Math. Phys.} {\bf 42}, 103-114; and {\it Ann. I.H.P.} {\bf 70}, 461-514].

\section*{Introduction}

Poincar\'e-Dulac normal forms (and their hamiltonian counterpart,
Birkhoff-Gus\-tavs\-son normal forms) are central to much of our understanding of local nonlinear dynamics. 

We refer to \cite{GMDE} for an exposition of the fundamental aspects 
-- and to \cite{EMSI} for an in-depth exposition -- of normal forms theory. Readable and reasonably sized introductions to the matter are  contained e.g. in \cite{Elp,Gle,GuH,IoA,Ver,Wal}; see also \cite{CGs}.
 
We want to discuss a point which has been investigated by several authors and which has an obvious interest both for theory and for concrete (including computational) applications, i.e. the possibility of further reducing a system which is already in normal form (NF in the following).

It is indeed well known that the NF classification is in general redundant for resonant systems, and different NFs can be conjugated. This fact, and the possibility of obtaining more favourable NFs by taking advantage of the freedom in the choice of the generators for the normalizing change of coordinates, were already remarked by Dulac in \cite{Dul}.

Among recent approaches to this problem, we recall
\cite{AFG,Bai,BaC,BaS,Bro2,BrT,CDD,ChK,KOW,vdM,Tak,Ush}. Many of them employ a similar set of ideas; among these, the use of the higher order analogues of the standard homological operator (already suggested in \cite{Tak}), and the Lie algebraic properties of the set of vector fields in normal form with respect to a given linear part (already suggested in \cite{Bro2}). Each approach has also, of course, its own features, and leads to similar but not identical results.
 
In \cite{LMP,IHP} I proposed an approach based on iteration of the standard Poincar\'e normalization procedure (the reduced normal forms thus obtained were therefore called Poincar\'e renormalized forms, or PRF for short); this is completely algorithmic and requires only to solve linear equations at each step.

The PRF algorithm uses only the grading by homogeneity degree of the algebra $\G_A$ of polynomial vector fields in normal form with respect to a given linear part $A$; it is thus quite general, but fails to take advantage of the Lie algebraic structure of $\G_A$. 
Not surprisingly, taking this Lie algebra structure into account can lead to a considerable simplification of the computations and of the resulting simplified normal form.

The main purpose of this note is indeed to conjugate the PRF approach with Lie algebraic considerations, so to obtain a procedure which takes advantage of the Lie algebra structure of $\G$ and keeps the computational simplicity of the PRF approach. Rather than aiming at the greater generality, we will focus on a structure which is non-generic, but relatively common in applications.

\bigskip

The discussion given in this note should also clarify some points related to standard PRF approach (as we discuss in the lines below); it should also make clear how much the PRF approach is related to Broer's ideas and to previous work along the same lines. Actually, the set of ideas employed here is essentially the same (higher order homological operators, Lie algebras filtrations) as in those works; they are blended here in a form which is suitable for concrete computational implementation.

It should be stressed that concrete computations in \cite{LMP,IHP} were performed by using this the Lie algebraic structure, although this point was not sufficiently stressed there; this could have caused some confusion, which the present note should hopefully dissipate. To make things worse and add occasions for confusion, in \cite{LMP,IHP} the term PRF was on the one hand precisely defined, but  on the other end also used to indicate generically reduced normal forms obtained by use of a sequence of Poincar\'e transformations, such as those based on the Lie algebra structure.
 
This can cause confusion due to the following fact: if we consider the sequence of Lie-Poincar\'e transformations prescribed by the PRF procedure, we obtain some reduced normal form $F_1$ (which is of course a PRF). If the sequence of Lie-Poincar\'e transformations is not taken in this order but according to a different scheme, e.g. to take advantage of the Lie algebra structure on $\G_A$ (as in considering the above mentioned example in \cite{LMP,IHP}), we obtain another reduced normal form $F_2$. In general $F_2 \not= F_1$, but moreover this $F_2$ can fail to be a PRF according to the proper definition; this happened for the main example in \cite{LMP,IHP} and will also be the case for the example B considered in detail below. 
 
\bigskip

The paper is organized as follows: in section 1 we recall the main facts about Poincar\'e-Dulac (standard) normal forms and fix notation; in section 2 we discuss two further reduction scheme, i.e. the ``generic'' PRF procedure and a modification of it which makes use of  the Lie algebraic structure of $\G_A$ and applies when this structure has favourable properties (this will be the LRF approach); in section 3 we briefly consider a three-dimensional example, with linear part corresponding to a center-focus, showing the simplicity of computations required by the LRF approach; in section 4 we consider a case where the structure of $\G_A$ is not the optimal one for use of the LRF approach, but still allows for its use. Finally, in section 5 we analize in full detail a simple two-dimensional example, giving a completely explicit description of the renormalizing transformations and renormalized forms up to order six, and compare the results obtained with the PRF and with the LRF approach. 

We also provide two appendices: in the first one we discuss Bruno's treatment of PRF and remark that his definition is not equivalent to the one given in \cite{LMP,IHP}, so that the example he considers in \cite{Brep,Bru2} -- and which falls in the case considered in section 5 -- does not apply to PRFs; in appendix B we briefly recall the main ideas put forward by Broer and Baider, and describe the Broer-Baider reduction procedure in the language employed in the present paper.

\subsection*{Acknowledgement}

This work was started in the Physics Department of Universit\'a di Roma, and completed in the Mathematics Department of Universit\'a di Milano. I would like to thank A. Degasperis and P. Santini (Roma) and D. Bambusi and L. Galgani (Milano) for their kindest hospitality. 
The support of ``Fondazione CARIPLO per la ricerca scientifica'' under the project {\it Teoria delle perturbazioni per sistemi con simmetria} is gratefully acknowledged.

\section{Standard normal forms}
\def\sn{1}

Let us first collect some basic formulas about (standard) normal forms.
We work in $\R^n$ with basis $\{ {\bf e}_1 , ... , {\bf e}_n \}$ and coordinates $(x^1 , ... , x^n )$; and consider a vector field $X$ in $\R^n$ having a zero in the origin. This is written in coordinates as $X = f^i (x) \pa_i$ (here and below, $\pa_i = \pa / \pa x^i$); we expand the vector function $f(x)$ in homogeneous terms as $f(x) = \sum_{k=0}^\infty f_k (x)$, with $f_k (ax) = a^{k+1} f_k (x)$. 
When considering coordinate expressions, we will write
$\V_k$ for the set of vector fields with components homogenoeus of order $(k+1)$ in the $x$. It is clear that under commutator we have $[ \V_k , \V_m ] \ss \V_{k+m}$.

We also consider the linearization $Ax$ of $f$ in the origin, given by
$A^i_{\, j} := (\pa f^i / \pa x^j ) (0)$. We write in general $f(x) = Ax + F(x)$, where $F(x)$ collects non\-lin\-ear terms only.

We denote by $ \s := \{ \la_1 , ... , \la_n \}$ the spectrum of $A$. 
If the $\la_i $ satisfy some relation
$$ \sum_{i=1}^n \mu_i \la_i := (\mu \cdot \la ) \ = \ \la_\a \eqno(1) $$
where $\a = 1,...,n$, $\mu_i$ are non-negative integers, and $|\mu | := \sum_i \mu_i \ge 2$, we say that $A$ is {\it resonant}.

A vector of the form $\vb = (x_1^{\mu_1} ... x_n^{\mu_n} ) \eb_\a$, with $\mu_i$ and $\a$ as in (1), is called a {\it resonant monomial vector}; the linear span of resonant monomial vector is the linear space of {\it resonant vectors}. When some ambiguity could arise, we specify these are {\it resonant with $A$}.

The Poincar\'e-Dulac theorem affirms that it is possible to find a
sequence of near-identity changes of coordinates such that in the new coordinates (denoted again by $x$) the vector field $X$ is expressed as $X = g^i (x) \pa_i$, with $g(x) = Ax + G(x)$, and with the nonlinear term $G(x)$ being the sum of resonant terms only. The change of coordinates defined by the whole sequence is in general only formal, but it is analytic (in some open neighbourhood of the origin) if zero lies out of the convex hull of $\s$ in the complex plane (Poincar\'e condition).

It is actually also possible to express the changes of coordinates as the time-one action of vector fields $H_k = h_k^i \pa_i$ (this has several theoretical and practical advantages \cite{BGG,Dep,MiL,Wal2}); in this case we speak of {\it Lie-Poincar\'e transformations}. 

When a nonlinear vector field $X = [A^i_{\, j} x^j + G^i (x) ] \pa_i$ in given coordinates satisfy the condition that the nonlinear part $G^i (x)$ is a resonant vector -- with respect to the linear part $A$ -- we say that $X$ is resonant (in the $x$ coordinates).

Notice that the resonance condition involves only the eigenvalues of $A$; if we decompose $A$ into a semisimple and a nilpotent part (mutually commuting), $A = A_s + A_n$, then only $A_s$ enters in the definition of resonant vectors. 

If $A$ is not semisimple, it is actually possible to refine ulteriorly the form to which $G (x)$ can be taken. Denoting $X_A = A^i_{\, j} x^j \pa_i$ the vector field associated to the linear part of $X$ (notice this is invariant under near-identity changes of coordinates), we denote by $\L_0$ the operator of commutation with $X_A$; that is, $\L_0 (Y) = [X_A , Y ]$. Obviously, $\L_0 : \V_k \to \V_k$. 

We can then force $G_k = G \cap \V_k$ to be in a complementary space to the range of $\L_0$ in each $\V_k$. 
If we introduce a scalar product\footnote{Choosing the Bargmann
scalar product \cite{Elp} will guarantee that $\L_0^+ (Y) = [X_{A^+} , X ]$, where $A^+$ is the adjoint of $A$. Another popular choice (standard scalar product) is described in \cite{GMDE}.} in $\V_k$ (and hence in $\V = \V_0 \cup \V_1 \cup ...$), then we can force $G \in \ker (\L_0^+ )$. 
When the condition $X_G \in \ker (\L_0^+ )$ is satisfied, we say that $X$ is in Poincar\'e-Dulac normal form (in the given coordinates). 

Notice that in general  $[X_A, X_{A^+}] \not= 0$, and thus the adjoint of the linear part of a vector field in normal form does not commute with the full vector field, but only with the nonlinear part. On the other side, if $[X_A , X_{A^+} ] = 0$  (which is equivalent to $[A,A^+] = 0$, i.e. to $A$ being a normal matrix), then $X_{A^+}$ is a symmetry of $X$.

Here we will {\it assume} that {\it $A$ is semisimple} $A = A_s$ and has been taken to {\it Jordan normal form}\footnote{The theory of normal forms is well developed without these assumptions, but
reduction encounters a number of substantial obstacles; see e.g.
\cite{ScW,Wal} for normal forms in the case where $A_n \not=0$ and/or $A$ is not in Jordan normal form.}, so that 
$$ A \ = \ {\rm diag} (\la_1 , ... , \la_n ) \ . \eqno(2) $$ 
Thus, in the present discussion, vector fields in normal form and resonant vector fields will be the same. We also have that $\L_0 = \L_0^+$ (provided we have chosen a suitable, e.g. the Bargmann, scalar product).

Resonant vector fields are thus identified by the condition
$$ \[ X_0 , X \] \ = \ 0 \eqno(3)$$
where $X_0$ is the linear part of $X$. In order to emphasize that we are
actually dealing with coordinate expressions and not with geometrical objects,
it is maybe better to deal with components of $X$ in the $x$ coordinates. It
is easy to see that if $X_f = f^i (x) \pa_i $ and $X_g = g^i (x) \pa_i$, then
$[X_f , X_g ] = X_h = h^i (x) \pa_i$, with 
$$ h (x) = \{ f (x) , g(x) \} \ := \ (f^j (x) \cdot \pa_j) g (x) - (g^j (x) \cdot \pa_j ) f^i (x) \ . \eqno(4)$$
The bracket $\{.,.\}$ is just expressing the commutator in terms of the components of vector fields in given coordinates. Relation (3) reads then 
$$ \{ [(Df)(0)] x , f(x) \} \ = \ 0 \ . \eqno(3') $$

The point is that once $f(x)$ is in normal form, we can still consider
near-identity Lie-Poincar\'e changes of coordinates taking it into a different normal form $\^f (x)$; in order to be guaranteed that $\^f (x)$ is still in NF, i.e. (3') is satisfied by $\^f$, the generator $h$ must be chosen to be itself resonant with $A = (Df)(0)$. Indeed, the set $\G_A$ of vector fields commuting with $X_A$ is obviously a Lie algebra.

We recall that if $h(x) = x + \sum_{k=1}^\infty h_k (x)$, the changes of coordinates given by the time-one flow of $X_h$ maps $X$ into $\^X = e^H X e^{-H}$. This can be explicitely computed by the classical 
Baker-Campbell-Haussdorff formula; with $\Theta (X) := [H,X]$, we have 
$$ \^X \ = \ \sum_{s=0}^\infty (1/s!) \Theta^s (X) \eqno(5)$$
which in terms of homogeneous components reads
$$ \^f_m \ = \ \sum_{s=0}^{[m/k]} \ {1 \over s!} \ \h^s (f_{m-sk}) \eqno(6)$$
where $[m/k]$ denotes the integer part of $(m/k)$ and $\h (f) = \{ h, f \}$.

It is convenient, for further discussion, to define the higher order homological operators (already considered by Takens \cite{Tak}) as $\L_k := \{ f_k , . \}$. 
We also denote, for later discussion, by $G = C(A)$ the centralizer of $A$ in the Lie algebra of $n$-dimensional matrices (with Lie operation the matrix commutator); a basis for this will be given by matrices $\{K_1 , ... , K_d \}$ (if $A \not= 0$, we can choose $K_1 = A$); notice that $d \le n$.

\section{Two non-equivalent further reduction \\ 
schemes: PRF and LRF}
\def\sn{2}

Having defined normal forms and obtained the formulas for the change of variables, we can wonder if one can change variables thus  transforming a normal form into another one, more convenient in some respect (in particular, having a smaller number of nonlinear terms, or a smaller number of low order ones). The answer is obviously yes, as already noted by Dulac \cite{Dul}, in the form and within the limits implied by (5), (6) above.

In recent works -- as mentioned in the Introduction --  I have proposed a general procedure (and algorithm,  requiring to solve only linear equations) for such a reduction \cite{LMP,IHP}.
Since this approach represents a direct extension of Poincar\'e procedure, the normal form (in general not unique) obtained in this way has been called ``Poincar\'e renormalized form'' (PRF), and we refer to it as the PRF approach.

On the other hand, in some cases it is also possible to proceed in a different way: one can use the Lie algebraic structure of the set of resonant vectors. We will call the reduced normal forms obtained in this way, ``Lie renormalized form'' (LRF), and we refer to the procedure  as the LRF approach.  

It should be mentioned that the relevance of Lie algebraic structures in normal forms theory was stressed (in his thesis) by Broer, who gave a very general and powerful reduction procedure (see appendix B). 

Our procedure will be less general and powerful, and apply only in fa\-vou\-ra\-ble cases (see below); but it is simpler and -- when applicable -- its practical implementation in completely explicit computations is elementary, as we also show by example.

\subsection{The PRF approach}

I will now recall the basic aspects of the PRF approach, developed in \cite{LMP,IHP}; the reader is referred to these works and to \cite{CGs} for further detail.

Let $W$ be the vector field (say already in standard NF, to avoid trivial steps) under consideration. Let us write it as $W = X_0 + \sum_{k=1}^\infty W_k$, with $X_0$ linear and $W_k$ homogeneous of degree $(k+1)$. For the sake of simplicity\footnote{Notice that the PRF can as well deal with cases where the linear part vanishes: now the standard homological operator is only the first in a chain of operators, and we can use the other ones for reduction. In practice, computations will be very difficult for $A=0$, unless some other constraint (e.g. symmetry properties) reduces the set of allowed nonlinear vector fields; see \cite{CLG} for PRF analysis of symmetric systems with $A=0$.} we will assume $X_0 \not= 0 $. 

Take the first nonzero $W_k$, say $W_{p_0}$; operating with transformations generated by $h_k^{(0)} \in \W_k^{(0)} := \ker (\L_0) \cap \V_k$ (successively for $k=1,2,...$) we can eliminate all terms in $[W_{p_0} , \W_k^{(0)} ]$, i.e. in the range of $\M_{p_0}$, defined as the restriction of the operator $\L_{p_0}$ to $\ker (\L_0)$. 

Let now $W_{p_1}$ be the first nonzero term among those with $k > p_0$ in the normal form obtained after the above transformations. 
Using transformations generated by $h_k^{(1)} \in \W_k^{(1)} := \ker (\L_0 ) \cap \ker (\L_{p_0} ) \cap \V_k$ (successively for $k=1,2,...$), we can eliminate all terms in $[W_{p_1} , \W_k^{(1)} ]$, i.e. in the range of $\M_{p_1}$, the restriction of $\L_{p_1}$ to $\ker (\L_0) \cap \ker (\L_{p_0})$. 

The process can obviously be continued indefinitely, until either all nonlinear terms of degree higher than $p_q$ are killed, or $\ker (\L_0) \cap \ker (\L_{p_0} ) \cap ... \cap \ker (\L_{p_q} )$ is empty.

Notice that in this way, due to the restriction to kernels of lower order ho\-mo\-lo\-gi\-cal operators (i.e. due to the use of $\M_p$ rather than $\L_p$), at each step we are not affecting the terms which have already been simplified; in facts at each step we stabilize new terms.

The generators $h_k^{(j)}$ are chosen as solutions to higher homological equations: if $f^{(j)}_k$ is the term of order $(k+1)$ after the first $j$ rounds of further normalization, and $\pi^{(j)}_k $ the projection from $\V_k$ to $\W^{(j)}_k$, this is given by 
$$ \pi^{(j)}_k \[ \sum_{s=0}^{[m/k]} \ {1 \over s!} \ \h^s (f_{m-sk}) \] \ = \ 0 \eqno(7)$$
where $\h = \{ h_k^{(j)} , . \}$; that is, see (5), by requiring that $\pi_k^{(j)} \( {\^W}_k^{(j)} \) = 0$.

Notice that the $h_k^{(j)}$ considered above are in general not uniquely defined: they are unique up to an element in $\ker (\M_{p_j})$, see above. Thus, the PRF is in general not unique.

The reduced normal form obtained according to this procedure will be called a {\it Poincar\'e renormalized form} (PRF) for the normal form  $W = W^{(0)}$.
A precise description of the spaces to which the terms non-eliminable in this way belong -- i.e. an abstract general description of PRFs -- and a more detailed discussion are given in \cite{CGs,LMP,IHP}.

\subsection{The LRF approach}

Let us now consider a different further normalization scheme.

Consider the set of vector fields in $\R^n$ which are in normal form with respect to the given linear part $A$, i.e. the set of $Y \in \V$ such that $[X_A , Y ] = 0$. It is obvious that these form a Lie algebra (the Lie operation being the standard commutator of vector fields); we denote this algebra by $\G$.
 
Let us recall a general characterization of vector fields in normal form relevant in this context \cite{CGs,Elp,IoA,Wal}. Consider the linear vector field $X_A$; we say that the differentiable function $\phi : \R^n \to \R$ is an invariant for $X_A$ if $X_A (\phi ) = 0$. 

Denote by $\I^* (A)$ the set of invariants for $X_A$ which are meromorphic (that is, can be expressed as a quotient of algebraic functions) in the $x$ coordinates; denote by $\I (A) \ss \I^* (A)$ the set of algebraic invariants for $X_A$, and by $\I_k (A) \ss \I (A)$ the set of algebraic invariants for $X_A$ which are functions homogeneous of degree $k+1$ in the $x$ variables.

Let $G = C(A)$ be the centralizer of $A$ in the algebra of $n$ dimensional matrices; let its Lie algebra be spanned by matrices $\{ K_1 , ... , K_d \}$ (we can always assume $K_1 = I$, and that $K_\a = A$ for some $\a$, provided $A \not= 0$; notice that $d \le n$). We denote by $X^{(\a)}$ the vector fields corresponding to these, i.e. given in the $x$ coordinates by $X^{(\a)} = (K_\a x)^i \pa_i$. 

Then the most general vector field $W$ in $\G$ can be written as
$$ W \ = \ \sum_{\a=1}^d \ \mu_\a (x) \ X^{(\a)}  \eqno(8)$$
where $\mu_\a (x) \in \I^* (A)$. In other words, $\G$ is contained in  a finitely generated module over $\I^* (A)$. 

Notice that the vector field $W$ must be algebraic in the $x$, and $X^{(\a)}$ are linear in $x$, so that functions $\mu_\a (x) \in \I^* (A)$ having poles of degree $d \ge 2$ in $x = 0$ cannot appear in (10). That is, only algebraic functions and functions with simple poles in the origin can appear in the actual normal form unfolding: $\G$ is not the full $G$-generated module over $\I^* (A)$. 

\bigskip\noindent
{\bf Example.} Let us briefly mention an example where indeed meromorphic functions of the invariants enter in the normal form unfolding. Consider systems in $\R^3$ with coordinates $(x,y,z)$; let the linear part be given by the diagonal matrix $ A = {\rm diag}(-1,1,2)$, so that $X_A = - x \pa_x + y \pa_y + 2 z \pa_z$.
This has two basic invariants, given by $\Psi_1 = xy$ and $\Psi_2 = x^2 z$. We take as $X^{(a)}$ the vectors
$ X^{(1)} = x \pa_x$, $X^{(2)} = y \pa_y$, and $X^{(3)} = z \pa_z$. 
It is immediate to check that $(\Psi_1^2 / \Psi_2) X^{(3)} = y^2 \pa_z$ and $(\Psi_2 / \Psi_1) X^{(2)} = xz \pa_y$ are polynomial and resonant with $X_A$. $\odot$
\bigskip

In several cases it happens that $\G$ has a more convenient structure, i.e. the $\mu_\a$ in (10) can actually be taken to be in $\I (A)$, and not just in $\I^* (A)$. In this case we say that all the vector fields in $\G$ are {\it quasi-linear}, or that we have a {\it quasi-linear normal form}. 
In particular, this is the case when $A$ admits only one basic invariant (see the examples below).

If the normal form is quasilinear, we have $\G \cap \V_{k+1} = \I_k (A) \otimes G$, and the analysis of the structure of $\G$ results to be particularly simple, as we now discuss.

Call $\X^*_\a$ the algebra spanned by vectors which are written as $X = s (x) X^{(\a)}$ with $s \in \I^* (A)$; call $\X_\a$ the algebra spanned by vectors as above with $s \in \I (A)$ (this is the module over $\I (A)$ generated by $X^{(\a)}$). 

As seen before, in general we have $ \X_1 \oplus ... \oplus \X_d \subseteq \G \subset \ \X^*_1 \oplus ... \oplus \X^*_d$, and in the (favourable) quasi-linear case we actually have $\G = \X_1  \oplus ... \oplus \X_d$.

Consider now the commutation relations between elements of the subalgebras $\X_\a^*$ and $\X_\b^*$; it is immediate to check that 
$$ \begin{array}{l}
\[ \mu_\a (\Psi ) X^{(\a)} , \s_\b (\Psi ) X^{(\b)} \] \ = \ \( \mu_\a (\Psi) \, (\pa \s_\b / \pa \psi_i ) \, X^{(\a)} (\psi_i) \) X^{(\b)} \\ 
\ \  \ - \ \( \s_\b (\Psi) \, (\pa \mu_\a / \pa \psi_i ) \, X^{(\b)} (\psi_i) \) X^{(\a)} \ + \ \( \mu_a (\Psi) \s_\b (\Psi) \) \, \[ X^{(\a )} , X^{(\b)} \] \ . \end{array} \eqno(9)$$
Notice that when $X^{(\b)} = X_A$, by definition $X^{(\b)} (\psi_i ) = 0$, and $[X^{(\a)} , X^{(\b)} ] = 0$; thus the corresponding subalgebra $\X_\b$ is always an abelian ideal in $\G$. 

\medskip\noindent
{\bf Remark.} Note also that, as obvious from the formula (9) above, the union of subalgebras $\X_{\a_1} \cup ... \cup \X_{\a_s}$ is a subalgebra in $\G$ if and only if $\{ X^{(\a_1 )} , ... , X^{(\a_s )} \}$ span a subalgebra in $G$. $\odot$
\medskip

It can happen that we are able to determine a sequence of subalgebras $\F_p \subseteq \G$, each of them being the union of $\X_\a$ subalgebras, such that $\F_0 = \G$ and 
$$ \[ \, \G \, , \, \F_p \, \] \ = \  \F_{p+1} \ ; \eqno(10) $$
if this terminates in zero we say that $\G$ has a {\it quasi-nilpotent} structure. Notice that the factor algebras $\Gamma_p := \F_p / \F_{p+1}$ are in general {\it not } abelian.

By the above remark, $\G$ can have a quasi-nilpotent structure only if $G$ is nilpotent. The chain of subalgebras $\F_p \subset \G$ can then be read off the descending central series $G_p$ of $G$; recall that the factor algebras $\gamma_p = G_p / G_{p+1}$ for this are abelian. 
The subalgebras $\Ga_p$ introduced above are therefore moduli over $\I (A)$ generated by abelian subalgebras $\gamma_p$ of $G$.

\bigskip

Assume now $\G$ is quasi-nilpotent. In this case we can first work with generators in $\Ga_1$ and simplify terms in $\Ga_1$ (e.g. by following the PRF algorithm within the set $\Ga_1$; this allows to work with more familiar projection and homological equations than if setting the problem in a completely Lie algebraic framework), then consider generators in $\Ga_2$ and simplify the corresponding terms being guaranteed that $\Ga_1$ terms are not changed, and so on. 

Notice that in this case we are -- roughly speaking --  just using the nilpotent structure of (the finite dimensional group) $G$, rather than the one of (the infinite dimensional algebra) $\G$.

Needless to say, this approach is particularly convenient when the $\Ga_p$ are generated by a single element of $G$.

The situation depicted above is met in applications: e.g., it applies to any nontrivial two-dimensional case and several three-dimensional ones \cite{prf2d}. More generally, it always applies when there is only one basic invariant.

In the following sections we will consider some simple examples where the LRF is easily computed, and it turns out to be definitely simpler than the PRF.

\section{Example I}
\def\sn{3}

Let us consider a system in $\R^3$ (we use coordinates $x,y,z$) with linear part given by 
$$ A \ = \ \pmatrix{0 & -1 & 0 \cr 1 & 0 & 0 \cr 0 & 0 & -1 \cr } $$
It is easy to see that this has only one basic invariant $\Psi := (x^2 + y^2 )$. As any meromorphic function of $\Psi$ is either algebraic or has a pole of degree $d \ge 2$ in the origin, we deduce that the most general vector field in normal form with respect to this linear part is
$$ W \ = \ X_A \ + \ \sum_{k=1}^\infty a_k X_k + b_k Y_k + c_k Z_k \eqno(11) $$
where (with $k \ge 0$)
$$ \cases{ 
X_k := \ \Psi^k \, (x \pa_x + y \pa_y ) & \cr
Y_k := \ \Psi^k \, (- y \pa_x + x \pa_y ) & \cr
Z_k := \ \Psi^k \, (z \pa_z ) & . \cr} $$
This form can also be easily deduced by explicit computation applying the definition of resonant vector field. Obviously, $X_A = Y_0 - Z_0$. We denote by  $\mu , \nu , \s$ the first $k \ge 1$ such that $a_k , b_k , c_k$ are nonzero. 

The $X_k,Y_k,Z_k$ satisfy the commutation relations
$$ \begin{array}{l}
\[ X_k , X_m \] = 2 (m-k) X_{k+m} \ , \ \[ Y_k , Y_m \] = 0 \ , \ \[ Z_k , Z_m \] = 0 \\
\[ X_k , Y_m \] = 2 m Y_{k+m} \ , \ \[ X_k , Z_m \] = 2m Z_{k+m} \ , \ \[ Y_k , Z_m \] = 0 \end{array} \eqno(12)$$

Denoting by $\X , \Y , \Z$ the algebras spanned by the $X_k $, the $Y_k$ and the $Z_k$, we have that $\G = \X \oplus \Y \oplus \Z$, and that $\Y \oplus \Z$ is an abelian ideal in $\G$. We can thus apply the LRF procedure discussed above. 

We first operate on $\X$, with generators also in $\X$ (thus $H_k = \a_k X_k$); in this way we can eliminate all terms except the $X_\mu$ and the $X_{2 \mu}$ ones, as implied by (12). In doing this we modify terms in $\Y \oplus \Z$.
 
Having performed this first step, we pass to consider the $\Y$ and $\Z$ terms, operating with generators in $\Y \oplus \Z$ (thus $H_k = \b_k Y_k + \gamma_k Z_k$). It is clear from (12) that we can eliminate all terms with $k > \mu$, but no lowest order ones. Thus we end up with a LRF given by 
$$ \^W \ = \ X_A \ + \ a_\mu X_\mu + \^a_{2 \mu} X_{2 \mu} \ + \ \sum_{k = \nu}^\mu \^b_k Y_k \ + \ \sum_{k = \s}^\mu \^c_k Z_k \ ; $$
the hat on constants mean that these are not the same as in the initial form (11), and obviously a sum with lower limit greater than the higher limit should just be meant as zero.

As shown by this example, the computations required for the determination of the general LRF are actually very simple.

\section{Example II}
\def\sn{4}

We want now to consider an example where the algebra $\G_A$ of resonant vector fields has not the optimal structure for LRF reduction, i.e. the $\Ga_p$ are not generated by a single element of $G$ (see section 2). 

We consider a system in $\R^4$ with linear part corresponding to two oscillators with nonzero and non-resonant frequencies $\a$ and $\b$, $|\a / \b | \not\in {\bf Q}$, i.e. in block notation 
$$ A \ = \ \pmatrix{ \a J & 0 \cr 0 & \b J \cr} \ \ \ , \ \  J = \pmatrix{0&-1\cr1&0\cr} \ ; $$
this $A$ has eigenvalues $\la = \pm i \a , \pm i \b$. 
We use coordinates $(x,y,z,w)$, so that
$$ X_A \ = \ \a \, ( -y \pa_x + x \pa_y ) \ + \ \b \, ( - w \pa_z + z \pa_w ) \ . $$
It is immediate to see that the system is simply resonant and admits two independent basic invariants, $ \psi_1 = x^2 + y^2$ and $\psi_2 = z^2 + w^2$. On the other hand, the linear space $\G_1$ of linear vector fields commuting with $X_A$ is spanned by $ X^{(\a)} = (M_\a x) \nabla$, with $\a = 1,...,4$ and we can choose the matrices $M_\a$ e.g. as (in block notation, with $I$ the two-dimensional identity matrix and $J$ as above)
$$ M_1 = \pmatrix{I & 0 \cr 0 & 0 \cr} \ , \ M_2 = \pmatrix{0&0\cr0&I\cr} \ , \ M_3 = \pmatrix{J&0\cr0&0\cr} \ , \ M_4 = \pmatrix{0&0\cr0&J\cr} $$
It is also immediate to check that $X^{(\a)} (\psi_i ) = 0$ except for $X^{(1)} (\psi_1 ) = 2 \psi_1$ and $X^{(2)} (\psi_2 ) = 2 \psi_2$.

We will write $ X^\a_{km} := \psi_1^k \psi_2^m X^{(\a)}$; these span $\G$, for $k,m$ non-negative integers. We also write, with the notation introduced above, $\G = \X_1 \oplus ... \oplus \X_4$; note that $\X_3 \oplus \X_4$ is an abelian subalgebra in $G$. 

The nontrivial commutation relations are given by 
$$ \begin{array}{l}
\[ X^1_{km} , X^1_{pq} \] \ = \ 2 (p-k) \, X^1_{k+p,m+q} \ \ , \ \ 
\[ X^2_{km} , X^2_{pq} \] \ = \ 2 (q-m) \, X^2_{k+p,m+q} \\ 
\[ X^1_{km} , X^2_{pq} \] \ = \ 2 p \, X^2_{k+p,m+q} \, - \, 2 m X^1_{k+p,m+q} \\ 
\[ X^1_{km} , X^\a_{pq} \]  =  2 p  X^\a_{k+p,m+q} \ \ , \ \ 
\[ X^2_{km} , X^\a_{pq} \]  =  2 q  X^\a_{k+p,m+q} \ \ 
(\a = 3,4) \end{array} $$
It is easy to see that there is no way to separate the algebra as the sum of one-dimensional moduli over $\I (A)$. Thus, the best we can do in the spirit of the LRF approach is to act first in the $\X_1 \oplus \X_2$ subalgebra. Once this has been reduced, say with lowest nonlinear term in $\G_\mu$, we can eliminate all terms in $(\X_3 \oplus \X_4) \cap \G_m$ for all $m > \mu$, as follows immediately from the commutation relations.

Note that the choice of terms to be eliminated in $\X_1 \oplus \X_2$ is to a large extent arbitrary, and correspondingly the LRF is not unique.

\section{Example III}
\def\sn{5}

We will now consider the LRF procedure for a simple system in full detail, i.e. aim at producing completely explicit formulas for the renormalized form and for the renormalizing transformation, thanks to the computational simplicity of the LRF procedure. We will also compare explicitely the LRF and PRF reductions.

Let us consider a two-dimensional system (in $\R^2$ with coordinates $x,y$) with linear part given by 
$$ A  =  \pmatrix{0&0\cr0&1\cr} $$
i.e. corresponding to the vector field $ y \pa_y $. 
We note immediately that here $A$ is diagonal, so that $A = A^+$. We have only one basic invariant $\psi (x,y) = x$.

\subsection{Normal forms}

It is easy to see that the kernel of $\L_0$ is spanned by the arrays of vector fields (with $k \ge 0$) 
$$ X_k  :=  x^{k+1} \, \pa_x \ \in \W_k \ \ {\rm and} \
\  Y_k  :=  x^k y \, \pa_y \ \in \W_k  $$
(with this notation the linear part considered here is given by $Y_0$). These vector fields satisfy the commutation relations
$$  
[ X_k , X_m ] =   (m-k) \, X_{k+m} \ 
, \ \ 
[ X_k , Y_m ] =   m \, Y_m \
, \ 
\ 
[ Y_k , Y_m ] =   0 \ .  \eqno(13) $$
We denote by $\X$ the algebra spanned by the $X_k's$, by $\Y$ the
algebra spanned by the $Y_k$'s; obviously $\G = \X \oplus \Y$. Note that $\Y$ is an abelian ideal in $G$. 

The (standard) normal form corresponding to the linear 
part considered in this section will thus be given by a vector field
$$ W  =  Y_0  +  \sum_{k=1}^\infty ( a_k X_k + b_k Y_k ) \eqno(14) $$
depending on the two infinite sequences of real constants $a_k , b_k$.

We will denote by $\mu$ (respectively, by $\nu$) the first $k \ge 1$ such that $a_k \not= 0$ (respectively, such that $b_k \not= 0$).
It results that for the sake of our present discussion, it is of interest to consider the case $\nu < \mu$ (for a full discussion of this system, whatever $\mu$ and $\nu$, see \cite{prf2d}). 
We will refer to the case $\nu = 1$ as nondegenerate, and to $\nu > 1$ as degenerate.

\subsection{The PRF reduction scheme}

We want now to consider the PRF corresponding to the linear part given by $A$. In the spirit of PRF, we should act on the NF (14) with 
Lie-Poincar\'e transformations generated by homogeneous functions $h_m \in \ker (\L_0) \cap V_m$. These will correspond to the action of vector fields of the form $ H_m =  \alpha X_m  +  \beta Y_m$. 

We have then to consider $\L_1$; this depends on the coefficients of the qua\-dra\-tic part $W_1$ of the vector field $W$, which we write as $W_1  =  a_1 X_1 + b_1 Y_1$. 

Under our assumption that $\nu < \mu$, necessarily $a_1 = 0$. In the nondegenerate case, $b_1 \not= 0$, while in the  degenerate one, $b_1 = 0$.
 
In the nondegenerate case we have $W_1 = b_1 Y_1$.
We notice that $[Y_1 , X_k ] = - Y_{k+1}$ and $[Y_1 , Y_k ] = 0$; therefore $\ker (\M_1 ) = \Y$. On the other hand, 
$\ran (\M_1)$ also is given by $\Y$, and $\ker (\M_1^+ ) = \X $.
In this case we also have to consider higher order parts of $W$; the first step of the PRF procedure can eliminate all terms in $\ran (\M_1 )$ and thus we will only consider terms in $\ker (\M_1^+)$. 

Let $\mu$ be as above, and let $W_\mu = a_\mu X_\mu$ (all the $Y_k$ parts with $k \ge 2$ can be eliminated, as just recalled). Now $\M_\mu$ is the restriction of $\L_\mu$ to $\ker (\M_1 ) = \ker (\L_0 ) \cap \ker (\L_1)$: indeed the $\L_m$ with $1 < m < \mu$ are zero and put no restriction. 
We have $ [X_\mu , Y_k ] = k Y_{k+\mu}$ and thus $\ker (\M_\mu ) = \{ 0 \} $: no further normalization is possible within the PRF scheme.
 
Thus the PRF is given in this case by  
$$ \^W  \ = \ Y_0 \, + \, b_1 \, Y_1 \, + \, 
\sum_{k=2}^\infty \, \^a_k X_k \ , \eqno(15) $$
where the hats on constants $\^a_k$ indicate that coefficients are in general not the same as those of the initial NF (14).
 
We anticipate that the LRF reduction scheme can give a finite dimensional normal form for this case, see below.

\bigskip

The previous discussion can easily be generalized to the degenerate  case where $1 < \nu < \mu$, with $\mu$ and $\nu$ defined above; notice that at least one of these has to exist and be finite, or the system would already be linear and thus trivial. 
Here the NF is 
$$ W \ = \ Y_0 \ + \ \sum_{k=\nu}^{\mu-1} b_k Y_\mu \ + \
\sum_{k=\nu}^\infty (a_k X_k + b_k Y_k ) \ . \eqno(16) $$ 
Now $\L_\nu (H_k) = b_\nu [ Y_\nu , \a_k X_k + \b_k Y_k ] = - \nu b_\nu \a_k Y_{\nu+k}$, and therefore we can eliminate all the $Y_{\nu+k}$ terms simply by choosing, with the same notation as before, $ \a_k = - \=b_{\nu+k} / (\nu b_\nu)$; we cannot eliminate any of the $X_k$ terms. Thus, the PRF in the degenerate case is
$$ \^W \ = \ Y_0 \ + \ b_\nu Y_\nu \ + \ \sum_{k=\mu}^\infty \=a_k X_k \ . \eqno(17) $$
Similarly to what happens for the nondegenerate case, the LRF procedure gives better results in this case.

\subsection{The LRF reduction scheme}

In the previous computations, we have followed the general PRF scheme for further normalizing the standard NF (14); this gave an infinite PRF in both the degenerate and nondegenerate cases.
 
However one can take advantage of the specific Lie algebraic structure of $\G = \X \oplus \Y$, embodied in (13), to obtain a more drastical reduction: indeed, one can obtain a reduction to a finite normal form (the Lie renormalized form), as we now discuss.
We use the same notation as in discussing the degenerate case above.
 
We first operate a sequence of normalizations with generators $h_k^{(a)} = \a_k X_k$, which we choose so as to eliminate higher order $X_k$ terms, i.e. $X_k$ for $k > \mu$ (as we know, this is not possible for $k = 2 \mu$). Notice this will change not only the (coefficients of the) $X_k$
terms, but the (coefficients of the) $Y_k$ terms as well; however, no terms of degree $k < \nu$ will be produced. 
 
In this way, we arrive at a partially reduced form (the tilde indicates that the coefficients are not the same as the initial ones, but not yet final)
$$ \=W \ = \ Y_0 \ + \ a_\mu X_\mu + \=a_{2 \mu} X_{2 \mu}
\ + \  \sum_{k=\nu}^\infty \=b_k Y_k \ . $$

Once this has been done, we pass to consider a second sequence of normalizations with generators $h_k^{(b)} = \b_k Y_k $. As $\Y$ is an ideal in $\G$, the $X_k$ terms are unaffected. On the other side, $\Y$ is abelian, and so only the $X_\mu$ and $X_{2 \mu}$ are actually active in these transformations: that is, we can only eliminate terms $Y_{\mu + 1}$ and higher (it is clear by the commutation relations that these can always be eliminated). 
 
In this way we arrive at the LRF: this is a NF depending on $(\mu - \nu + 3)$ constants (recall we assumed $\mu > \nu$), of the form
$$ \^W \ = \ Y_0  \ + \ a_\mu X_\mu + \^a_{2 \mu} X_{2 \mu} \ + \ 
\sum_{k=\nu}^\mu \^b_k Y_k \ . \eqno(18) $$
It is also clear by this discussion that actually  $\^b_k = {\widetilde b}_k$, $\^a_{2 \mu } = {\widetilde a}_{2 \mu}$. Note that the number of constants ($\mu - \nu + 3$) agrees with that computed by Bruno \cite{Brus}.

It should be stressed that this LRF is {\it not } a PRF, as can be checked by comparing this with (15) and (17) above, or comparing the definition of PRF. 

Indeed, with the notation employed in \cite{LMP,IHP}, the spaces $F^{(k)}_k := F^{(k)} \cap \W_k$ with $\nu < k \le \mu$ reduce to multiples of $X_k$. Here we have therefore $W_k \not\in F^{(k)}_k$ for $\nu < k \le \mu$, and thus the LRF cannot be a PRF.

\subsection{Explicit reduction (nondegenerate case)}

The reductions described in previous subsections can be explicitely performed; detailed computations are reported in \cite{prf2d}, while here we just give results. We write the normal form $W$ in the form (14) and consider further normalization up to order six. 

Let us first consider the PRF reduction.
In the nondegenerate case we will take, for the sake of simplicity, $\b_k = 0$; we choose $\a_1 = - b_2 / b_1$, $\a_2 = (b_2^2 - b_1 b_3 )/ b_1^2$, $\a_3 = - (2 b_2^3 - 3 b_1 b_2 b_3 + b_1^2 b_4) / b_1^3$, $\a_4 = (9 b_2^4 - 18 b_1 b_2^2
b_3 + 3 b_1^2 b_3^2 + 8 b_1^2 b_2
 b_4 - 2 b_1^3 b_5) / (2 b_1^4)$. In this
way we obtain 
 $$ \begin{array}{rl}
 \~W^{(5)} \ = &\ Y_0 \ + \ b_1 \, Y_1 \
+ \ a_2 \, X_2 \ + \ 
 [ a_3 - a_2 b_2 / b_1 ] \, X_3 \ + \\
  & + \ [ a_4 - 2
a_3 b_2 / b_1 + a_2 b_2^2 / b_1^2 ] \, X_4 \ + \\
  & + \ [ a_5 - 3 a_4 b_2 /
b_1 + 4 a_3 b_2^2 / b_1^2 + \\
  & \ \ \ \ - a_3 b_3 / b_1 - 2 a_2 b_2 b_3 /
b_1^2 + a_2 b_4 / b_1 ] \, X_5
 \ + \ O(6) \ . \end{array} $$
\bigskip

Let us now pass to consider the LRF reduction, and perform detailed computations according to the LRF scheme for the nondegenerate case; that is, we first take care of the $X_k$ terms (up to $k=5$), and then of the $Y_k$ ones. 

With a transformation $h_1 = \a_1 X_1$, the $W_3$ term reads
$$ \=W_3 \ = \ \[ a_3 + a_2 \a_1 \] X_3  \ + \ \[ b_3 + 2 b_2 \a_1 + b_1 \a_1^2 \] Y_3 \ . $$
We disregard the $Y_3$ term and choose $\a_1$ so to eliminate the $X_3$ term, i.e. $\a_1 = - a_3 / a_2$. 
After computing the effect of this on higher order terms, we could perform a transformation with generator $h_2 = \a_2 X_2$. However, we know that there will be no way to eliminate the $X_4$ term, so we set $\a_2 = 0$. We perform a transformation with generator $h_3 = \a_3 X_3$.  With this, the $W_5$ term reads
$$ \begin{array}{rl}
\=W_5 \ =& \ [ 2 a_3^3 /a_2^2 - 3 a_3 a_4 / a_2 + a_5 - a_2 \a_3 ] X_5 \ + \\ 
 & + \ [ a_3^4 b_1 / a_2^4 - 4 a_3^3 b_2 / a_2^3 + 
   6 a_3^2 b_3 / a_2^2 - 4 a_3 b_4 / a_2 + \\ 
 & + \ \ \ b_5 - 2 a_3 b_1 \a_3 / a_2 + 2 b_2 \a_3 ] Y_5 \ . 
\end{array} $$
Again we only aim at eliminating the $X_5$ term, and thus we choose
$ \a_3  =  [( 2 a_3^3 - 3 a_2 a_3 a_4 + a_2^2\ a_5) / (a_2^3)]$.
We will be satisfied with this order of normalization for the $X_k$ terms, and take now care of the $Y_k$ ones.

We first operate a transformation with generator $h_1 = \b_1 Y_1$; we get 
$$ \=W_3 \ = \ \[ a_3^2 b_1 / a_2^2 - 2 a_3 b_2 / a_2 + b_3 -  a_2
\b_1 \] \ Y_3 \ . $$
 By choosing 
$ \b_1  =  [( a_3^2 b_1 - 2 a_2 a_3 b_2 + a_2^2 b_3) / (a_2^3)]$
we eliminate this. We compute the effect on higher order term, and then consider a transformation with generator $h_2 = \b_2 Y_2$; with these, we have
$$ \begin{array}{rl}
\=W_4 \ =& \ \[ a_4 - a_3^2 / a_2 \] \ X_4  \ + \\  
 & \ + \ (1/a_2^3) \ [ a_3^3 b_1 + 3 a_2 a_3^2 b_2 - 
        3 a_2 a_3 (a_4 b_1 + a_2 b_3) +  \\
 & \ \ +  a_2^2 (a_5 b_1 + a_2 (b_4 - 2 a_2 \b_2)) \, ] \ Y_4 \ . 
\end{array} $$
We want to eliminate the $Y_4$ term, and thus we choose
$ \b_2  =  (1/2 a_2^4) \, ( a_3^3 b_1 - 3 a_2 a_3 a_4 b_1 + a_2^2 a_5 b_1 + 3 a_2 a_3^2 b_2 - 3 a_2^2 a_3 b_3 + a_2^3 b_4)$. 
Again we take into account the effect of this on higher order terms, and pass to consider a transformation with generator $h_4 = \b_4 Y_4$; we get 
$$ \begin{array}{rl}
\=W_5 \ =& \ [ - 2 a_3^4 b_1 / a_2^4 + 
   5 a_3^2 a_4 b_1 / a_2^3 -  2 a_3 a_5 b_1 / a_2^2 - 
   2 a_3^3 b_2 / a_2^3 - 4 a_3 a_4 b_2 / a_2^2 + \\
 & \ + 2 a_5 b_2 / a_2 + 7 a_3^2 b_3 / a_2^2 - 
   a_4 b_3 / a_2 - 4 a_3 b_4 / a_2 + b_5 - 3 a_2 \b_3 ] \ Y_5 
\end{array} $$
which can be eliminated by choosing 
$ \b_3 =  
-  (1 /( 3 a_2^5 )) \, (2 a_3^4 b_1 - 5 a_2 a_3^2 a_4 b_1 + 
 2 a_2^2 a_3 a_5 b_1 + 2 a_2 a_3^3 b_2 + 
4 a_2^2 a_3 a_4 b_2 - 2 a_2^3 a_5 b_2 - 
 7 a_2^2 a_3^2 b_3 + a_2^3 a_4 b_3 + 
 4 a_2^3 a_3 b_4 - a_2^4 b_5 ) $.

Summarizing, and having taken into account all higher order effects (up to order six), we have reached the LRF
$$ \^W \ = \ Y_0  +  b_1 Y_1  +  a_2 X_2  +  [ b_2 - (a_3 b_1 / a_2 ) ] Y_2  +  [a_4 - (a_3^2 / a_2 ) ] X_4  + \ O(6) $$

\vfill\eject

\section*{Appendix A. \\ Bruno alternative definition and example.}

In his recent book \cite{Bru2} (and before this is \cite{Brep}), A.D. Bruno has dealt with PRFs. Unfortunately he reports a definition for PRFs which is not equivalent to the one contained in \cite{LMP,IHP} (and reported above), so that his comments on PRFs can improve confusion rather than clarifying the issue, especially since the difference in definitions is not made clear (nor mentioned).

The definition of PRFs given in Bruno's works (which I take from section V.22 of \cite{Bru2}) is as follows, once translated to the notation used in the present paper (the definition given in the first review \cite{Brep} is slightly different, but the difference is unessential here).

We consider the vector power series $f_k$; let $F_m = \sum_{k=0}^m f_k$; the operators $\Psi_k$ are then defined as $\Psi_m (H) = (H^j \pa_j ) F_m - (F_m^j \pa_j ) H$; in the present notations, we have
$$ \Psi_m \ = \ - \, \sum_{k=0}^m \, \L_k \ . \eqno(A.1)$$
Bruno considers then the adjoint $\Psi_m^*$ of the operators $\Psi_m$, and declares that PRFs have the property that 
$$ \Psi_{k-1}^* (F_k) \ = \ 0 \ ; \eqno(A.2)$$
this is not equivalent to my original definition. 
In facts, this condition implies that each term $G_k$ is in $\ker (\L_p^+ )$ for all the $p < k$; in my definition \cite{LMP,IHP} this applies to $\M_p^+$ rather than $\L_p^+$, which obviously makes a substantial difference.

Bruno considers then an example, i.e.  the system
$$ {\dot x} \ = \ x^3 \ \ , \ \ {\dot y} \ = \ y (1 + x + x^2 ) \ ; \eqno(A.3)$$ 
according to \cite{Bru2} the PRF for this is given by ($a_2 \not= 0$)
$$ {\dot x} \ = \ a_2 x^3 + \a x^5 \ \ , \ \ {\dot y} \ = \ y (1 + \b x) \eqno(A.4)$$ 
with no higher order terms. In \cite{Brep} the PRF is instead  claimed to be
$$ {\dot x} \ = \ x^3  \ \ , \ \ {\dot y} \ = \ y (1 + x) \ , \eqno(A.5)$$
again with no higher order term (the difference between these two expressions is never mentioned, nor explained, in both of \cite{Brep,Bru2}). In both cases, Bruno does not explain how these expressions are obtained.
However, both (A.4) and (A.5) are obviously different from the real PRF, see the example discussed in detail in section 5. 
Bruno proves then that (A.3) cannot be conjugated to (A.4) or (A.5); we have seen that these are not PRFs and thus his arguments do not concern PRFs (in the sense considered in \cite{LMP,IHP} and here).

\vfill\eject

\section*{Appendix B.\\ General Lie algebraic structure of normal forms}

A general approach employing the Lie algebraic structure of nonlinear vector fields resonant with a given linear one was proposed by Broer and Takens \cite{Bro2,BrT}, and further developed by other authors.
The Broer procedure is always applicable, and is able produce a unique reduced normal form \cite{Bai,BaC}. Unfortunately, implementing it in practice seems to be quite hard, and this approach has been applied only to system of very low dimension.
On the other side, the central ideas put forward by Broer are at the basis of many approaches to the problem of further reduction of normal forms (in particular the one discussed in this note) and I will describe them the way I understand them.

Given a matrix $A$, we denote by $\G$ the Lie algebra of vector fields in normal form with respect to $A$. One can consider the descending central series (DCS) of $\G$ \cite{Kir,NaS}. We recall this is defined by $\G_0 = \G$ and $\G_{k+1} = [ \G , \G_k ]$; as well known the factor algebras $\Ga_k = \G_k / \G_{k+1}$ are abelian. It is then possible to eliminate terms by inner automorphisms of $\G$ (that is, by acting on $\G$ with vector fields in $\G$) proceeding along $\G_k$, i.e. filtering the Lie algebra $\G$. 
This is relevant to our problem since Lie-Poincar\'e transformations reduce to inner operations on $\G$.

This approach is, of course, particularly convenient when $\G$ is nilpotent; actually if we consider nonlinear vector fields resonant with a given $A$, i.e. $\G^* := \G \backslash (\G \cap \V_0)$, and $\G^*_k$ its DCS, we have $\G^*_k \cap \V_m = \emptyset$ for $m<k$, and thus $\G^*$ is nilpotent of infinite order.
We can reduce to the more familiar case of algebras nilpotent of finite order $N$ by considering only the projections of $\G_k$ and $\Ga_k$ to the linear space $\V^N \subset \V$ spanned by nonlinear vector fields homogeneous of degree not higher than $N$.

Let us now briefly describe the implementation of the BRF procedure. We consider the sequence of abelian factor algebras $\Ga_k = \G_{k+1} / G_k$, and denote by $\chi_k$ the projection from $\G$ to $\Ga_k$. 

We will produce a sequence of vector fields $W^{(j)}_*$, each being the result of the first $j$ steps of further normalization. Let $\eta_j$ be the operator on $\G$ defined by $\eta_j (H) = e^H W^{(j)}_* e^{-H}$; let $\pi_j$ be the operator of projection from $\G$ to the range of $\eta_j$. We can reduce the normal form $W = W^{(0)}_*$ as follows. 

As the first step, consider a $H^{(0)} \in \G$ and require that $ \~W = W^{(1)}_* \ := \eta_0 (H)$ is such that $ \pi_0 [ \chi_0 \eta_0 (H^{(0)}) ] = 0 $. This determines (non uniquely) $H^{(0)}$, and produces a $W^{(1)}_*$. 

At further steps, we have the same setting; the ``homological equations'' on Lie algebras to be solved at each step will be 
$$ \pi_j \, \[ \, \chi_j \( \eta_j (H^{(j)}) \) \] \ = \ 0 \ ; \eqno(B.1)$$
this determines $H^{(j)}$. Each $W^{(j+1)}_*$ is then determined as 
$$ W^{(j+1)}_* \ = \ \eta_j (H^{(j)} ) \ := \ \exp [H^{(j)}] \, \exp [W^{(j)}_* ] \, \exp [ - H^{(j)} ] \ . \eqno(B.2) $$

\vfill\eject

\end{document}